**A note on the optimum allocation of resources to follow up unit nonrespondents in probability surveys**

By SM Tam[1,2], A Holmberg[2] and S Wang[3]

[1.] National Institute of Applied Statistical Research, University of Wollongong, Australia. 1, Northfields Ave, Wollongong, NSW 2522, Australia. Stattam@gmail.com

[2.] Australian Bureau of Statistics, Canberra, Australia. 45, Benjamin Way, Belconnen, ACT 2615, Australia. Anders.Holmberg@abs.gov.au

[3.] Australian Bureau of Statistics, Canberra, Australia. 45, Benjamin Way, Belconnen, ACT 2615, Australia. Summer.Wang@abs.gov.au>

Abstract: Common practice to address nonresponse in probability surveys in National Statistical Offices is to follow up every nonrespondent with a view to lifting response rates. As response rate is an insufficient indicator of data quality, it is argued that one should follow up nonrespondents with a view to reducing the mean squared error (MSE) of the estimator of the variable of interest. In this paper, we propose a method to allocate the nonresponse follow-up resources in such a way as to minimise the MSE under a quasi-randomisation framework. An example to illustrate the method using the 2018/19 Rural Environment and Agricultural Commodities Survey from the Australian Bureau of Statistics is provided.

Acknowledgement

We would like to thank the referees and the Associate Editor for their helpful comments on an earlier version of this paper. The views expressed in this paper are those of the authors and do not necessarily reflect those of the Australian Bureau of Statistics.

## 1 Introduction

Nonresponse is unavoidable and has become an increasingly challenging issue worldwide for survey practitioners. Item or unit nonresponse, unless properly adjusted for in estimation, will usually have a negative impact on survey data quality (Groves 2006). An extensive literature on nonresponse adjustment exists in journals and books and will not be repeated here – see for example, Bethlehem (1998), Groves and Couper (1998), Hedlin (2020), Kim and Kim (2014), Kim and Shao (2014), Little (1986), Little and Rubin (2019), Oh and Scheuren (1983), Särndal et al.(1992), Särndal and Lundström (2005), Schouten et al.(2011), Sikov (2018) to name just a few.

Recent literature on survey design addresses the nonresponse focusses on adaptive designs (see, for example, Groves and Heeringa 2006). Adaptive designs use survey para data and auxiliary data to guide changes to the procedures during data collection to minimise the unit cost of collection whilst addressing nonresponse. Quoting Beaumontet al. (2014), Neusy et. al. (2022) argued that adaptive collection procedures, such as call prioritization, cannot reduce the nonresponse bias to a greater extent than a proper nonresponse weight adjustment. We agree with this view. Accordingly, we will not consider adaptive designs further in this paper.

In this paper, we examine a particular aspect of mitigating the effects of unit nonresponse. We address a classic trade-off for official statistics between cost and quality when collecting data, namely, the optimal deployment of resources earmarked by the survey statistician for nonresponse follow-up (NFU), to meet the NFU budget and maximise data quality. As far as we are aware, this topic has not been addressed in the literature. For example, whilst Neusy et.al. (2022) provided a method to determine the NFU sample size, its main aim is to ensure that the NFU budget is not exceeded, and did not, as a twin objective, seek out to maximise the quality of the statistics as a result of the NFU resources deployment.

Currently, a common NFU practice in probability surveys, including those carried out in national statistical offices, is to follow up every nonrespondent with the ultimate aim of maximising the response rate. It is well known that response rates are not necessarily good indicators of survey quality (Curtin et al. 2000; Groves et al. 2004; Groves 2006; Groves and Peytcheva 2006). Indeed, Schouten et al. (2009) gave an example of a Dutch survey in which an increase the response rate of 12.5% points led to an increase of bias of between 0.1 to 0.5% points.

It is therefore clear that increasing response rate in NFU is not the main game. The main game should be to reduce the mean squared error of the estimator of the finite population parameter of interest, subject to the constraint of the NFU budget. This is the main purpose of this paper.

The approach we advocate for the allocation of NFU resources to maximise data quality is not the same as taking a random sample of the non-respondents for follow up – see for example, Elliott et al. (2000), Hansen and Hurwitz (1946), Särndal et. al. (1992, 566) and Neusy et. al. (2022). Our approach allows all nonrespondents to be followed up, but the number of visits to a nonrespondent is determined by where it is located in the Response Homogeneity Group (RHG) (Särndal et. al. 1992, 578) and the unit cost for NFU in that Group. A challenge with subsampling of the nonrespondents approach is that it requires the assumption of full response in the follow-up phase which is not easy to achieve. On the other hand, the quasi-randomisation framework used in the estimation (Oh and Scheuren, 1983) accepts nonresponse as an outcome in the follow-up phase and accounted for it by using weights computed by the inverse of the estimated propensity scores. We use this framework in the paper for nonresponse mitigation.

This weighting approach, on the other hand, requires the assumption that (1) every nonrespondent has a non-zero probability to respond and (2) the response mechanism is missing at random. If there is "power of compulsion" enshrined in the statistics legislation, and if the national statistical office uses this power to conduct the survey, there is a good chance that the first assumption is fulfilled. Where the model for the missing-not-at-random (MNAR) nonresponse mechanism is correctly specified, Kim and Morikawa (2022), using an empirical likelihood method and a result from Pfeffermann and Sverchkov (1999) - refer to Note 4 below – showed that the inverse propensity weights (IPWs) can be calibrated to adjust for nonresponse bias and benchmark constraints. In a lecture presented to the 2022 Summer School in Ottawa,

Canada, on Modern Methods in Survey Sampling, Professor Jae Kim extended the idea of Kim and Morikawa (2022) to multiple MNAR nonresponse mechanisms, thus reducing the reliance of a particular nonresponse model for the Kim and Morikawa (2022) approach to work. He showed that as long as one of the multiple NMAR models is correctly specified, the estimator is consistent.

In this paper, we assume that the nonresponse mechanism is missing at random within the Response Homogenous Groups (RHGs) (Särndal et. al., 1992, 578). Using this assumption, we propose a method, similar to the approach used for Neyman allocation, to allocate the nonresponse follow-up resources with a view to minimising the mean squared error of the estimator of the population total.

## 2 Notation and some well-known results

We assume we have a probability sample, $s$, drawn from a population $U$ of size $N$ with known first order, $\pi_i$, and second order, $\pi_{ij}$, inclusion probabilities for $i, j \in s,$ with $\pi_{ii} = \pi_i.$ We define the sample weights, $d_i$, by $d_i = \frac{1}{\pi_i}.$ Associated with each unit in $U$ is a target variable of interest, $y_i$, and a vector of auxiliary variables, $\boldsymbol{x}_i$. We are interested in estimating the population total $T = \sum_{i \in U} y_i.$ Due to unit nonresponse, we only have a responding sample, $s_r, s_r \subset s.$ Following Oh and Scheuren (1983), we model the nonresponse process using a quasi-randomisation (QR) model, where the responding units are considered to be selected in two stages with the first stage selected from $U$ by probability sampling characterised by known $\pi_i$'$s$ and $\pi_{ij}$'$s$, and the second stage selected from $s$ by a missing-at-random nonresponse process with the probability of responding, also referred to as an (unobserved) propensity score, $\rho_i, i \in s,$ defined by $\rho_i = \Pr(\delta_i = 1 | \boldsymbol{x}_i, i \in s : \boldsymbol{\alpha})$, where $\delta_i$ is the response indicator and $\boldsymbol{\alpha}$ is

vector of unknown parameters. In practice, $\rho_i$ is estimated by $\hat{\rho}_i = \Pr(\delta_i = 1 | \boldsymbol{x}_i, i \in s : \hat{\boldsymbol{\alpha}})$ where $\hat{\alpha}$ is a consistent estimator of $\boldsymbol{\alpha}$ by assuming a functional form of the probability distribution of $\rho$, e.g. $\frac{\rho}{1-\rho}$ is modelled by logistic regression; or by using non-parametric methods, such as random forest (Breiman 2001).

When using weights to mitigate the effects of nonresponse during estimation, it is customary in national statistical offices to use RHGs for nonresponse adjustment. A RHG is one where the estimated $\rho_i$ 's of its group members are the same, i.e missing at random within the RHGs. In reality, they are not the same but similar estimated $\rho_i$ 's can be grouped together to form nonresponse adjustment "cells". Little (1986) described this approach as response propensity stratification.

Suppose we partition the sample units by their estimated propensity scores into $H$ RHGs, with the $h^{th}$ group denoted by $s_h$. Then $s = s_1 \cup .. \cup s_h .. \cup s_H$, and let $n_h$ and $m_h$ denote the size of the sample and the responding units in $s_h$ respectively. Let $\rho_{hi}$ denote the response propensity for the $i^{th}$ unit in the $h^{th}$ RHG. Under the assumption that $\rho_{hi} = \rho_h$, where $\rho_h$ is a constant for every $i \in s_h$, we note that $n_h^{'} = E(\sum_{i \in s_h} \frac{\delta_{hi}}{\rho_{hi}} | s)$. Using the actual sample size, $n_h$, as a plug-in estimate of $n_h^{'}$, we have $n_h \simeq \sum_{i \in s_h} \frac{\delta_{hi}}{\rho_{hi}} = \frac{m_h}{\rho_h}$ thus giving an estimate of $\rho_h$, $\hat{\rho}_h = \frac{m_h}{n_h}$, which is just the response rate for the $h^{th}$ RHG. Likewise, let $\rho_{ij} = \Pr(\delta_i = \delta_j = 1 | x_i, x_j | s)$. Assuming that $Cov(\delta_{hi}, \delta_{hj})$ is a constant and $Cov(\delta_{hi}, \delta_{h'j}) = 0$ for $i \in h, j \in h'$ and $h \neq h'$, we can likewise show that $\hat{\rho}_{ij} = \frac{m_h(m_h - 1)}{n_h(n_h - 1)}$ for $i, j \in h, i \neq j$ and $\hat{\rho}_{ij} = \frac{m_h}{n_h}\frac{m_{h'}}{n_{h'}}$ for $i \in h, j \in h'$ and $h \neq h'$. In the sequel, we let $\pi_{hi}, \pi_{hij}, d_{hi}$ and $\boldsymbol{x}_{hi}$ denote the corresponding first order, second order inclusion probabilities, the weight and the vector of auxiliary variables for unit $i$ in the $h^{th}$ RHG.

Assuming $\Pr(m_h \leq 1)$ is negligible for $h = 1, ..., H$, we have the following well known results for nonresponse adjustment using RHGs. In the sequel, "unbiased" is defined in the QR sense, i.e. the QR expectation of the estimator equals to the quantity being estimated.

Known result 1 (Särndal et.al. 1992, 581). The QR IPW estimator of $T$ defined by $\hat{T}_1 = \sum_{h=1}^{H}\sum_{i \in s_h} \frac{\delta_{hi} d_{hi} y_{hi}}{\hat{\rho}_h}$ is asymptotically unbiased for $T$. In addition:

$$V\left(\hat{T}_1\right) = \sum_{i \in U}\sum_{j \in U} \Delta_{ij} \breve{y}_i \breve{y}_j + E_p E_{\boldsymbol{m}} \left(\sum_{h=1}^{H} n_h^2 \frac{1 - \rho_h}{m_h} S_{1h}^2 \mid s\right) \triangleq V_1\left(\hat{T}_1\right) + V_2\left(\hat{T}_1\right)$$

where $\Delta_{ij} = \pi_{ij} - \pi_i \pi_j$, $\breve{y}_i = \frac{y_i}{\pi_i}$, $E_{\boldsymbol{m}}(. \mid s)$ is the expectation with respect to the sampling distribution of $\boldsymbol{m} = (m_1, .., m_h, ..., m_H)^T$ given $s$, $S_{1h}^2$ is the variance of $\breve{y}_{hi}$ in $s_h$ and $\breve{y}_{hi} = \frac{y_{hi}}{\pi_{hi}}$.

Finally, an approximately unbiased estimator of the variance, $V\left(\hat{T}_1\right)$, is given by:

$$\hat{V}\left(\hat{T}_1\right) = \sum_{i \in U}\sum_{j \in U} \frac{\Delta_{ij}}{\hat{\rho}_{ij}} \delta_i \breve{y}_i \delta_j \breve{y}_j + \sum_{h=1}^{H} n_h^2 \frac{1 - \hat{\rho}_h}{m_h} S_{1hr}^2$$

where $S_{1hr}^2$ is the variance of $\breve{y}_{hi}$ in $s_{hr}$, i.e. the responding sample in $s_h$. In addition, $\hat{\rho}_{ij}$ is defined below.

(a) $\hat{\rho}_{ii} = \hat{\rho}_h = \frac{m_h}{n_h}$ for $i \in s_h$;

(b) $\hat{\rho}_{ij} = \frac{m_h}{n_h}\frac{m_h - 1}{n_h - 1}$ for $i, j \in s_h$ and $i \neq j$;

(c) $\hat{\rho}_{ij} = \frac{m_h}{n_h}\frac{m_{h'}}{n_{h'}}$ for $i \in s_h$ and $j \in s_{h'}$.

If the generalised regression estimator (GREG) is used instead of Horvitz-Thompson estimator in the QR IPW estimator of Result 1, we have the following result.

<u>Known result 2</u> (Kim and Kim 2014; Särndal and Lundström 2005, 52; Särndal et.al. 1992, 584). The QR IPW estimator of $T$ defined by

$\hat{T}_2 = \sum_{h=1}^{H}\sum_{i \in s_h} \frac{\delta_{hi} d_{hi} g_{hi} y_{hi}}{\hat{\rho}_h}$, is asymptotically unbiased for $T$, where

$g_{hi} = 1 + (\sum_{i \in U} \boldsymbol{x}_i - \sum_{h=1}^{H}\sum_{i \in s_h} \frac{\delta_{hi} d_{hi}}{\hat{\rho}_h} \boldsymbol{x}_{hi})^T (\sum_{h=1}^{H}\sum_{i \in s_h} \frac{\delta_{hi} d_{hi}}{\hat{\rho}_h} c_{hi} \boldsymbol{x}_{hi} \boldsymbol{x}_{hi}^T)^{-1} (c_{hi} \boldsymbol{x}_{hi})$ and $c_{hi}$'s are specified constants (often set as 1, see Särndal and Lundström 2005, 48) related to error structure of the estimation model underpinning GREG. Furthermore, analogous to Result 1, we have the following approximate variance:

$$V(\hat{T}_2) = \sum_{i \in U}\sum_{j \in U} \Delta_{ij} \breve{e}_i \breve{e}_j + E_p E_{\boldsymbol{m}} (\sum_{h=1}^{H} n_h^2 \frac{1 - \rho_h}{m_h} S_{2h}^2 \mid s) \triangleq V_1(\hat{T}_2) + V_2(\hat{T}_2)$$

where $\breve{e}_i = \frac{\hat{e}_i}{\pi_i}$ , $\hat{e}_i = y_i - \boldsymbol{x}_i^T (\sum_{h=1}^{H}\sum_{i\in s_h}\frac{\delta_{hi}d_{hi}}{\rho_{hi}}c_{hi}\boldsymbol{x}_{hi}\boldsymbol{x}_{hi}^T)^{-1}(\sum_{h=1}^{H}\sum_{i\in s_h}\frac{\delta_{hi}d_{hi}}{\rho_{hi}}c_{hi}\boldsymbol{x}_{hi}y_{hi})$, $S_{2h}^2$ is the variance of $\breve{e}_{hi}$ in $s_h$, $\breve{e}_{hi} = \frac{\hat{e}_{hi}}{\pi_{hi}}$ , $\hat{e}_{hi} = y_{hi} - \boldsymbol{x}_{hi}^T (\sum_{h=1}^{H}\sum_{i\in s_h}\frac{\delta_{hi}d_{hi}}{\rho_{hi}}c_{hi}\boldsymbol{x}_{hi}\boldsymbol{x}_{hi}^T)^{-1}(\sum_{h=1}^{H}\sum_{i\in s_h}\frac{\delta_{hi}d_{hi}}{\rho_{hi}}c_{hi}\boldsymbol{x}_{hi}y_{hi})$ Finally, an approximately unbiased estimator of the variance, $V\left(\hat{T}_2\right)$, is given by:

$$\hat{V}\left(\hat{T}_2\right) = \sum_{i\in U}\sum_{j\in U}\frac{\Delta_{ij}}{\hat{\rho}_{ij}}\delta_i\breve{e}_i\delta_j\breve{e}_j + \sum_{h=1}^{H} n_h^2\frac{1-\hat{\rho}_h}{m_h}S_{2hr}^2$$

where $\hat{\rho}_{ij}$ is given in Result 1 and $S_{2hr}^2$ is the variance of $\breve{e}_{hi}$ in $s_{hr}$ and

$$\hat{\breve{e}}_{hi} = \frac{y_{hi}}{\pi_{hi}} - \frac{\boldsymbol{x}_{hi}^T}{\pi_{hi}}(\sum_{h=1}^{H}\sum_{i\in s_h}\frac{\delta_{hi}d_{hi}}{\hat{\rho}_h}c_{hi}\boldsymbol{x}_{hi}\boldsymbol{x}_{hi}^T)^{-1}(\sum_{h=1}^{H}\sum_{i\in s_h}\frac{\delta_{hi}d_{hi}}{\hat{\rho}_h}c_{hi}\boldsymbol{x}_{hi}y_{hi})$$

If instead of weighting, imputation is used to mitigate the effect of nonresponse, we have the following result.

Known result 3 (Beaumont 2005). The QR Calibrated Imputation Estimator defined by $\hat{T}_3 = \sum_{i\in s} d_i g_i\{y_i + (1-\delta_i)y^*\}$ is asymptotically unbiased, where

$$y_i^* = \hat{\mu}_i + \frac{d_i g_i}{u_i}\left\{\sum_{i\in s}\frac{(1-\delta_i)(d_i g_i)^2}{u_i}\right\}^{-1}\left\{\sum_{h=1}^{H}\sum_{i\in s_h}\frac{1-\hat{\rho}_{hi}}{\hat{\rho}_{hi}}\delta_{hi}d_{hi}g_{hi}(y_{hi}-\hat{\mu}_{hi})\right\},$$

$$\hat{\mu}_i = \boldsymbol{x}_i^T(\sum_{h=1}^{H}\sum_{i\in s}\frac{\delta_{hi}c_{hi}\boldsymbol{d}_{hi}\boldsymbol{x}_{hi}\boldsymbol{x}_{hi}^T}{\hat{\rho}_{hi}})^{-1}(\sum_{h=1}^{H}\sum_{i\in s}\frac{\delta_{hi}c_{hi}\boldsymbol{d}_{hi}\boldsymbol{x}_{hi}y_i}{\hat{\rho}_{hi}}),$$

$\hat{\mu}_{hi} = \boldsymbol{x}_{hi}^T(\sum_{h=1}^{H}\sum_{i\in s}\frac{\delta_{hi}c_{hi}\boldsymbol{d}_{hi}\boldsymbol{x}_{hi}\boldsymbol{x}_{hi}^T}{\hat{\rho}_{hi}})^{-1}(\sum_{h=1}^{H}\sum_{i\in s}\frac{\delta_{hi}c_{hi}\boldsymbol{d}_{hi}\boldsymbol{x}_{hi}y_i}{\hat{\rho}_{hi}})$ and $u_i = \frac{1}{\hat{\sigma}_i^2}$ or $\frac{d_i g_i}{\hat{\sigma}_i^2}$ . Furthermore, the variance of $\hat{T}_3$ is given by:

$$V\left(\hat{T}_3\right)=\sum_{i\in U}\sum_{j\in U}\Delta_{ij}\breve{e}_i\breve{e}_j+E_pE_{\boldsymbol{m}}(\sum_{h=1}^{H}n_h^2\frac{1-\rho_h}{m_h}S_{3h}^2\mid s)\triangleq V_1\left(\hat{T}_3\right)+V_2\left(\hat{T}_3\right)$$

where $\breve{e}_i=\dfrac{y_i-\hat{\mu}_i}{\pi_i}$, $S_{3h}^2$ is the variance of $\breve{e}_{hi}$ in $s_h$ and $\breve{e}_{hi}=\dfrac{y_{hi}-\hat{\mu}_{hi}}{\pi_{hi}}$ Finally, an asymptotically unbiased estimator of the variance, $V\left(\hat{T}_3\right)$, is given by:

$$\hat{V}\left(\hat{T}_3\right)=\sum_{i\in U}\sum_{j\in U}\frac{\Delta_{ij}}{\hat{\rho}_{ij}}\delta_i\breve{e}_i\delta_i\breve{e}_j+\sum_{h=1}^{H}n_h^2\frac{1-\hat{\rho}_h}{m_h}S_{3hr}^2$$

where $\hat{\rho}_{ij}$ is given in Result 1 and $S_{3hr}^2$ is the variance of $\breve{e}_{hi}$ in $s_{hr}$.

<u>Note 1</u>: The approach in Result 3 is described by Beaumont (2005) as calibrated imputation because $\hat{T}_3$ is calibrated to an asymptotically QR unbiased estimator of the population total based on $\hat{\mu}_i$'s (refer equation 3.1 of Beaumont 2005). Also, some authors e.g. Beaumont (2005) use the more theoretically correct expression, $\breve{e}_i^*=g_i\breve{e}_i$, in lieu of $\breve{e}_i$ in Results 2 and 3. However, Särndal and Lundström (2005, 37) pointed out in practice, $\breve{e}_i^*\simeq\breve{e}_i$, so numerically there is little difference in using $\breve{e}_i$ instead.

<u>Note 2</u>: Oh and Scheuren (1983) proposed the variance in Result 1 be conditional on $\boldsymbol{m}$. We shall follow this approach in the sequel. As a result, we can drop the $E_{\boldsymbol{m}}$ operator in $V_2\left(\hat{T}_j\right), j=1,2,3$ in Results 1 to 3 above and replace it by $V_2(\hat{T}_j\mid\boldsymbol{m})$

<u>Note 3</u>: From the similarity in the formulas for the variance of the three estimators in Results 1 to 3, we can develop a generic approach for the optimal allocation of follow-up resources in the next Section. The approach outlined in this paper can be considered as the NFU resources allocation counterpart of Neyman allocation (Särndal et. al. 1992, 106) for sampling resources.

Note 4: As mentioned earlier, if the propensity score model is missing not at random, i.e. $\rho_i(\boldsymbol{x}_i, y_i;\boldsymbol{\alpha}) = \Pr(\delta_i = 1 | \boldsymbol{x}_i, y_i, i \in s : \boldsymbol{\alpha}),$ one can use $\frac{1}{\tilde{\rho}_i(\boldsymbol{x}_i | \boldsymbol{\alpha})}$ as the weight instead, where $\tilde{\rho}_i(\boldsymbol{x}_i | \boldsymbol{\alpha}) = \int \rho_i(\boldsymbol{x}_i, y_i | \boldsymbol{\alpha}) f(y_i | \boldsymbol{x}_i) dy_i$ and Pfeffermann and Sverchkov (1999) showed that $\frac{1}{\tilde{\rho}_i(\boldsymbol{x}_i | \boldsymbol{\alpha})} = E\left(\frac{1}{\rho_i(\boldsymbol{x}_i, y_i;\boldsymbol{\alpha})} | \boldsymbol{x}_i, \delta_i = 1\right)$. By assuming the function form of $f(y_i | \boldsymbol{x}_i, \delta_i = 1;\boldsymbol{\beta})$, one can use the respondent data to obtain a consistent estimate of $\boldsymbol{\beta}$ to compute

$$E\left(\frac{1}{\rho_i(\boldsymbol{x}_i, y_i;\boldsymbol{\alpha})} | \boldsymbol{x}_i, \delta_i = 1\right).$$

## 3 Optimum allocation of nonresponse follow-up resources

Current practice of NFU in probability surveys is often based on an approach trying to convert as many nonrespondents as possible with a view to increasing the response rate. As argued in Curtin et al. (2000), Groves et al. (2004); Groves (2006); Groves and Peytcheva (2006), response rate is an insufficient indicator of survey data quality and an increase in response rate does not necessarily lead to a reduction in nonresponse bias. In this paper, a strategic approach to NFU is proposed. Under this approach, the objective for NFU, under a fixed follow-up budget, is to minimise the mean squared error of the corresponding estimates. Given Results 1 to 3 above and note 2, this is equivalent to minimising $V_2\left(\hat{T}_j | \boldsymbol{m}\right), j = 1, 2, 3,$ i.e. the nonresponse variance, as the sampling variance, $V_1\left(\hat{T}_j\right)$, is not numerically affected in a NFU setting.

We now model the $k_h^{th}$ conversion event for the $i^{th}$ nonrespondent in the $h^{th}$ RHG as a random (Bernoulli) variable, $\lambda_{khi}$, with $\Pr(\lambda_{khi} = 1) = \rho_h$ and $\Pr(\lambda_{khi} = 0) = 1 - \rho_h$ , where $\lambda_{khi} = 1$ if the conversion is successful. Note that, by the definition of RHGs, every nonrespondent in the same RHG has the same probability of a successful conversion. Moreover, we assume that the events $\lambda_{khi}$ and $\lambda_{lhj}$ are independent, i.e. $\lambda_{khi} \perp \lambda_{lhj}$ where $k \neq l$ and $i \neq j$ and $h = 1,...,H; i, j \in U$. Whilst the assumptions $\lambda_{khi} \perp \lambda_{lhj}, i \neq j$ seem reasonable, particularly when the data collector for the $i^{th}$ nonrespondent is different to that of the $j^{th}$ nonrespondent, the assumptions $\lambda_{khi} \perp \lambda_{lhi}, k \neq l$ would hold only if the NFU visits are "passive" i.e. inability to establish contacts in all previous visit until the visit resulting in a successful conversion. On the other hand, if the visits are non-passive, it is likely that these visits will impact the nonrespondent's decision to participate in the survey during the current visit, e.g. persuasion effect as $k$ grows, or alternatively, a "no" in the previous visit is the ultimate position of the nonrespondent. In this case, the assumption $\lambda_{khi} \perp \lambda_{lhi}, k \neq l$ is violated. Nevertheless, the Bernoulli assumption, as a working assumption, is helpful to guide the allocation of NFU resources and benign to the estimation process which uses, amongst other things, the actual number of nonrespondents converted in the estimation process, including the calculation of the estimated variance.

Under the Bernoulli model, the probability of conversion, after $k_h$ visits for a nonrespondent in the $h^{th}$ RHG , is $k_h\rho_h$. Here, we restrict $k_h \leq \frac{1}{\rho_h}$, as a nonrespondent cannot be converted more than once, i.e. $k_h\rho_h > 1$ does not make sense. With $n_h - m_h$ nonrespondents in the $h^{th}$ stratum, the expected number of additional respondents converted, after $k_h$ attempts, is $k_h\rho_h(n_h - m_h)$. With respondents increased from $m_h$ to $m_h + k_h\rho_h(n_h - m_h)$, whilst the sampling variance term, i.e. $V_1\left(\hat{T}_j\right)$, of $V\left(\hat{T}_j\right), j = 1,2,3$ remains the same, we have:

$$V_2\left(\hat{T}_j \mid \boldsymbol{m}\right) = \sum_{h=1}^{H} n_h^2 \frac{1-\rho_h}{m_h + k_h\rho_h(n_h - m_h)} S_{jh}^2$$

$$\hat{V}_2\left(\hat{T}_j \mid \boldsymbol{m}\right) = \sum_{h=1}^{H} n_h^2 \frac{1-\hat{\rho}_h}{m_h + k_h\hat{\rho}_h(n_h - m_h)} S_{jhr}^2 \qquad (1)$$

where $S_{jh}^2$ and $S_{jhr}^2$ are defined in Results 1 to 3 respectively, depending on the type of weighting adjustment for nonresponse used.

On the other hand, the cost $c_h$ of following up $n_h - m_h$ nonrespondents in the $h^{th}$ RHG will be $u_h k_h (n_h - m_h)$, where $u_h$ is the unit cost per visit to convert, and the cost of following up all nonrespondents across all RHGs is

$$C = \sum_{h=1}^{H} u_h k_h (n_h - m_h). \qquad (2)$$

Thus, the strategic approach to NFU becomes finding $k_h$'s such that (1) is minimised subject to:

(a) $C \le C_0$ for a pre-determined $C_0$, where $C$ is defined in (2); (b) $k_h \le \frac{1}{\hat{\rho}_h}$; (c) $k_h$ is an integer; and (d) $k_h \le k_0$, where $k_0$ is another pre-determined constant to ensure reasonable respondent load. As there is no closed form for the solution, we have to find it using a solver. We used Excel Solver as illustrated in the next Section. It is good at finding solutions for problems with multiple inputs subject to multiple constraints. Note that the solutions are intuitively similar to the Neyman allocation for strata sample sizes - RHGs with high nonresponse variance is allocated with large follow-up resources, but tempered (in a non-linear way) by the unit cost of NFU effort.This is to boost the responding sample size needed to reduce the nonresponse variance for the relevant RHGs.

## 4 A numerical example

We tested the methods outlined in the previous Section through an empirical study, using data from the 2018/19 Rural Environment and Agricultural Commodities Survey (REACS) of the Australian Bureau of Statistics (ABS). Conducted annually, the REACS releases statistics on the production of agricultural commodities including cereal and broadacre crops, fruit and vegetables and livestock on Australian farms. The REACS sample comprised a stratified simple random sample of Australian farms. Like many ABS probability surveys, REACS was confronted with the challenge of declining response rates. For this empirical study, the variable of interest, i.e. the production of sheep, is illustrated. The sample for this variable of interest has 4,696 units (i.e. farms) in total, of which, 3,525 units were continuing units and 1,171 were new units, i.e. units first rotated into the 2018/19 survey.

To predict the response propensity scores for continuing units, the Random Forest (RF) with regression trees algorithm (Breiman 2001) was used. The RF algorithm is used in this paper instead of logistic regression models for predicting propensity scores for a number of reasons. For example, unlike logistic regression models, RF does not require the assumption of linearity or additivity in the modelling (Lee et al. 2010). Equally important, the automatic interaction detection inherent in RF algorithms provides a straightforward way to account for and allow easy interpretations of interactions between auxiliary data and the propensity to respond (Phipps and Toth 2014; Buskirk and Kolenikov 2015). For the RF algorithm used in this example, variables such as state, industry, size, statistical significance of the unit and such paradata as number of calls on the unit in previous survey cycles, number of reminder letters previously sent etc. were included in the model as predictors. Finally, the RF algorithm used in this example to predict propensity scores was chosen based on 10-fold cross validation (Hastie et al. 2009, 181), and has the smallest "out of sample" mis-classification rate in survey participation amongst the candidate RFs with different number of variable splits or trees.

For new units rotated into the 2018/19 REACS, the RF algorithm cannot be used to predict the propensity scores because para data from previous survey cycles for these units do not exist. Instead, their propensity scores were imputed by the average of the propensity scores of the continuing units considered to be their k Nearest Neighbours (kNN) (Hastie et al.2009, 463). To run the kNN algorithm, four variables from the 2018/19 REACS including state, natural resources management region, industry and size were used to calculate the "distance" metric.

Because they are categorical variables, the Gower's Distance (Gower 1971) was used as the distance metric to find the nearest neighbours.

What value of k should be used? The optimal k should be one that gives the most accurate out-of-sample predictions. To find the optimal k, 10-fold cross validation was again applied to the (training) dataset comprising the estimated propensity scores of the continuing units. The root mean squared error of the predicted propensity scores was used to measure and compare the accuracy of twelve kNN models for k = 1,2,…,12. As can be seen from Figure 1, whilst the 12NN model appears to be the most accurate, its RMSE is, however, not significantly different from that of the 5NN model. As the training dataset itself is skewed towards RHGs with high propensity scores, a small k is preferred, lest the KNN induced propensity scores of the new units are skewed by high propensity scores donors. Based on these considerations, the optimal k is taken to be 5 and the 5NN model was chosen to impute the propensity scores for the new units.

Figure 1. Estimated out-of-sample error rate by different value of k in kNN models

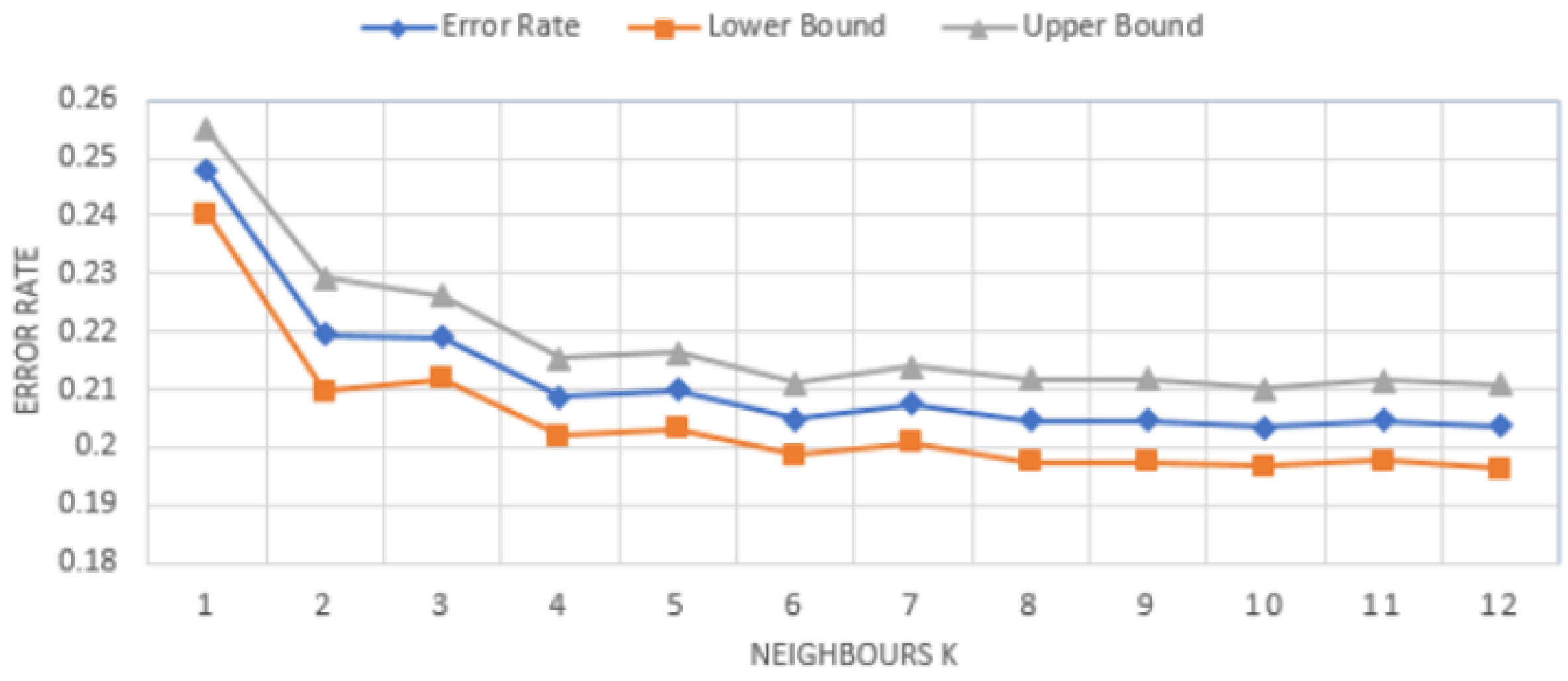

Finally, in REACS 2018/19, as the sample was stratified, and the stratified estimator was recast as a GREG estimator (Särndal and Lundström 2005, 37), so that the QR IPW estimator of Result 2, $\hat{T}_2$, is used in the numerical example.

Table 1 gives the relevant information with respect to the estimator of Result 2. In this example, we set $C_0 = \$30,000$ and $k_0 = 6$. We used the complimentary Excel Solver to solve the problem by creating an objective "cell" in Excel (i.e. equation (1)), a 10x1 vector of $k_h$ input "cells", three 10x1 vector of constraint "cells" (i.e. constraints (b) to (d) in Section 3 above), and a total cost constraint "cell" (i.e. constraint (a)). When running the Solver, Excel allows the 10x1 vector of $k_h$ input "cells" to vary, subject to meeting the requirements of the constraint cells, in order to find the solutions to minimise the objective cell. Table 1 gives the relevant data to compute the optimal solutions, and $\hat{V}_2\left(\hat{T}_2 \mid \boldsymbol{m}\right)$. The optimum solutions are given in Table 2.

Table 1: REACS 2018/19 information on $\hat{T}_2$ for sheep and $\hat{V}_2\left(\hat{T}_2 \mid \boldsymbol{m}\right)$ with no NFU

| RHG | Propensity Score Range | $n_h$ | $\hat{\rho}_h$ | $\hat{S}^2_{jhr}$ ($10^7$) | $m_h$ | $u_h$ (\$) | $\hat{V}_2\left(\hat{T}_2 \mid \boldsymbol{m}\right)$ ($10^9$) |
|---|---|---|---|---|---|---|---|
| 1 | (0,0.1] | 23 | 0.09 | 2.30 | 2 | 20 | 5.55 |
| 2 | (0.1,0.2] | 23 | 0.17 | 3.32 | 4 | 20 | 3.64 |
| 3 | (0.2,0.3] | 37 | 0.27 | 1.58 | 10 | 20 | 1.58 |
| 4 | (0.3,0.4] | 79 | 0.35 | 2.50 | 28 | 20 | 3.60 |
| 5 | (0.4,0.5] | 117 | 0.45 | 2.72 | 53 | 20 | 3.84 |
| 6 | (0.5,0.6] | 178 | 0.55 | 4.36 | 98 | 20 | 6.33 |
| 7 | (0.6,0.7] | 285 | 0.65 | 8.49 | 186 | 20 | 12.87 |
| 8 | (0.7,0.8] | 257 | 0.75 | 6.51 | 193 | 20 | 5.55 |
| 9 | (0.8,0.9] | 504 | 0.85 | 7.90 | 429 | 20 | 6.96 |
| 10 | (0.9,1) | 3193 | 0.95 | 4.69 | 3034 | 20 | 7.85 |
| Total | | 4696 | | | 4037 | | 57.77 |

In this example, the investment of about \$19.6k (see Table 2) for NFU is cost effective, as it reduces the nonresponse variance, $\hat{V}_2\left(\hat{T}_2 \mid \boldsymbol{m}\right)$, from 57.77 billion units (Table 1) where there is no NFU to about 38.21 billion units (Table 2) by restricting the number of follow-up visits to the values of $k_h$ as shown in Table 2. We can also see from Table 2 that there is an increase of 508 respondents with the optimal allocation of NFU resources, lifting the pre-follow up response rate of 86.0% to a post- follow up response rate of 97%. On the other hand, using the same Excel Solver, we can show that if only, say, \$10k is available for follow-up, the optimal $k_h$'s are $k_1 = 2, k_2 = 3, k_9 = k_{10} = 0$, and $k_i = 1,$ for $i = 3,4,5,6,7,8.$ This would incur a total investment of \$9.6k; gives a nonresponse variance of 41.9 billion units, and an increase of 225 respondents with a post-follow up response rate of only 91%.

With a sampling variance of 41.3x$10^9$ units for sheep production, the use of optimal NFU allocation reduces the mean squared error from 99.11 x$10^9$ (with non NFU follow-up) to 79.55 x$10^9$ units, or a reduction by 19.7% and 10.4% in the mean squared error and root mean squared error respectively.

It is also instructive to compare the optimal allocation strategy with the common practice, where there is no differentiation in the number of visits across the RHGs, i.e. $k_h = k_0, h = 1,\ldots,10.$ From (2), $k_0 = C_o / \sum_{h=1}^{H} u_h (n_h - m_h) = \$30{,}000 / \{\$20 * (4{,}696 - 4{,}037)\} \simeq 2$. As compared with the optimal allocation, we see that the common practice in this example does not put enough resources for the first 2 RHGs, and too much for the last 5 RHGs. The result is that the nonresponse variance of 39.84 billion units under the constant $k_0$ approach is higher than the optimum nonresponse variance of 38.21 billion units. Even though the reduction in mean squared error is 3.9%, the common practice approach costs $10.4k more than the optimal NFU allocation approach of $19.6k. However, with the number of additional respondents at 596 under the common approach, it gives a response rate of 99%. This is an example showing the response rate of 99% can give a misleading impression of providing a higher quality estimate of the number of sheep produced than the one with a response rate 97% with optimal NFU allocation.

Table 2: Optimal $k_h$ and $\hat{V}_2\left(\hat{T}_2 \mid \boldsymbol{m}\right)$ for $C_0 = \$30k$ and $k_0 = 6$

| RHG | RP Score Range | $k_h$ | Additional respondents* - $k_h\hat{\rho}_h(n_h - m_h)$ | RHG follow up cost ($) - $u_h k_h(n_h - m_h)$ | $\hat{V}_2\left(\hat{T}_2 \mid \boldsymbol{m}\right)$ ($10^9$) |
|---|---|---|---|---|---|
| 1 | (0,0.1] | 6 | 11 | $2,520 | 0.86 |
| 2 | (0.1,0.2] | 5 | 17 | $1,900 | 0.93 |
| 3 | (0.2,0.3] | 2 | 15 | $1,080 | 0.70 |
| 4 | (0.3,0.4] | 2 | 36 | $2,040 | 1.75 |
| 5 | (0.4,0.5] | 2 | 58 | $2,560 | 2.08 |
| 6 | (0.5,0.6] | 1 | 44 | $1,600 | 4.37 |
| 7 | (0.6,0.7] | 1 | 64 | $1,980 | 9.56 |
| 8 | (0.7,0.8] | 1 | 48 | $1,280 | 4.44 |
| 9 | (0.8,0.9] | 1 | 64 | $1,500 | 6.06 |
| 10 | (0.9,1) | 1 | 151 | $3,180 | 7.45 |
| Total | - | - | 508 | $19,640 | 38.21 |

*Rounded to the nearest integer

If one is interested in looking at the return on the investment of NFU resources, as measured by the reduction of $\hat{V}_2\left(\hat{T}_2 \mid \boldsymbol{m}\right)$ per unit NFU cost, one can use different values of $C_0$ to provide the requisite information as shown in Table 3. We observe that the highest return on the NFU investment is to spend about $10k for NFU, where 1091x$10^3$ units of variance is reduced per unit cost. However, the strategy to choose the highest return on NFU resources for NFU is only sensible if the nonresponse variance of 41.91 billion units is acceptable.

Table 3: Optimal $\hat{V}_2\left(\hat{T}_2 \mid \boldsymbol{m}\right)$ and Response Rate for $k_0 = 6$ and different values of $C_0$

| Cost ($) (not greater than) | $\hat{V}_2\left(\hat{T}_2 \mid \boldsymbol{m}\right)(10^9)$ | Response rate (%) | Actual cost ($) | Reduction in variance $(10^3)$ per additional unit cost |
|---|---|---|---|---|
| 5,000 | 47.26 | 88 | 4,600 | - |
| 7,500 | 44.42 | 89 | 7,380 | 1,022 |
| 10,000 | 41.91 | 91 | 9,680 | 1,091 |
| 12,500 | 40.56 | 92 | 12,200 | 536 |
| 15,000 | 39.24 | 93 | 14,900 | 489 |
| 17,500 | 39.02 | 96 | 17,500 | 85 |
| 20,000 | 38.21 | 97 | 19,640 | 379 |
| 20,250 | 38.21 | 97 | 19,640 | - |

## 5 Conclusion

In this paper, we provide a method to allocate NFU resources in such a way as to minimise the nonresponse variance (and thus the MSE) of the estimator of a variable of interest in probability surveys. This method requires the use of a quasi-randomisation framework and an inverse propensity weight to adjust for nonresponse.

In the numerical example, we used Random Forest to estimate the response propensity scores for the continuing units and applied the kNN algorithm to impute the propensity scores for the new units first rotated into the 2018/19 REACS. In both cases, selection of the RF algorithm, or the value of k in the kNN algorithm, was based on 10 fold cross validation, and the RF algorithm/k in the kNN algorithm with the least out-of-sample classification/prediction error was chosen. When compared with the common practice, the numerical example shows that there is a further reduction (about 4%) in the nonresponse variance $\hat{V}_2\left(\hat{T}_j \mid \boldsymbol{m}\right)$ by adopting the optimum allocation of NFU resources with only two third of the cost. The numerical example also shows that response rate is an inadequate indicator of data quality and the common NFU practice is not the most cost effective.

We note that official surveys are multi-purpose and optimising the allocation of NFU resources with respect to one target variable may not result in the optimal allocation of this resources to the other variables. This issue is also faced by survey statisticians in using Neyman allocation. An approach commonly used in official surveys to address this issue is to allocate the sample sizes with reference to the highest priority variable. Another way is to define objective functions for optimisation which address the multivariate situation. This is discussed in Holmberg (2002, 2003) and Holmberg et al. (2003). The same approaches can be used to allocate NFU resources in multiple purpose surveys.

The important messages of this paper are (1) reinforcing the well-known fact that allocating significant resources to boost response rates will not necessarily lead to better quality statistics; (2) in order to achieve better statistical outcomes, the NFU resource allocation should instead aim at minimising the variance from nonresponse; and (3) optimal allocation of NFU resources will result in the most cost effective NFU practice.